\documentclass{article}
\usepackage{geometry} 
\usepackage{amsmath}
\usepackage{amssymb}
\usepackage{accents}
\usepackage{latexsym}
\usepackage{multirow}
\usepackage{subcaption}
\usepackage{authblk}
\usepackage{hyperref}
\usepackage{graphicx}
\usepackage{bmpsize}
\usepackage[square,numbers]{natbib}  
\usepackage{color} 
\hypersetup{
setpagesize=false,
 bookmarksnumbered=true,%
 bookmarksopen=true,%
 colorlinks=true,%
 linkcolor=blue,
 citecolor=red,
}
\makeatletter
 
  \@addtoreset{equation}{section}
\makeatother

\title{Quintessential Inflation in Logarithmic Cartan $F(R)$ Gravity}

\author[1,2,3]{Tomohiro Inagaki }
\author[3]{Masahiko Taniguchi}

\affil[1]{Information Media Center, Hiroshima University, Higashi-Hiroshima, 739-8521, Japan}
\affil[2]{Core of Research for the Energetic Universe, Hiroshima University, Higashi-Hiroshima, 739-8526, Japan}
\affil[3]{Graduate School of Advanced Science and Engineering, Hiroshima University, Higashi-Hiroshima, 739-8526, Japan }

\date {} 
\begin{document}
\maketitle

\begin{abstract}
We investigate the quintessential inflation in the logarithmic Cartan $F(R)$ gravity.
A small logarithmic modification of the general relativity
has the potential to introduce both inflation and dark energy. 
We evaluate the time evolution of the Universe such as inflation, reheating, and dark energy.
The parameters in the model are fixed to introduce the inflation and the dark energy scales.
We show that the CMB fluctuations induced by the inflation are consistent with the current observations.
In the reheating process, it is possible to achieve the reheating temperature required for nucleosynthesis in Big Bang scenario.
It can be seen that by choosing an appropriate value for the scalaron field after reheating, the scalaron field again dominates the energy of the Universe and causes the current accelerating expansion as dark energy.
\end{abstract}

\section{Introduction}
The early and late-time accelerating expansion of the Universe is an interesting issue in cosmology.
The early-time expansion, i.e., inflation, is necessary to solve the horizon and flatness problems~\cite{Sato:1980yn,Guth:1980zm}.
A lot of theories explaining inflation have been proposed by extending the gravity sector or adding new matter with
minimal or non-minimal coupling to gravity~\cite{Bassett:2005xm,Maurya:2023soo,Santos:2023nhs,Taghavi:2023ptn}.
The scalar-tensor theory is a theory that places the dominant energy in the early Universe on the scalar degrees of freedom. 
If the potential of the scalar field has a flat plateau, the slow-roll scenario can be adapted.
In this case, the potential energy induces the accelerating expansion of the Universe.

Current observations such as type Ia supernovae~\cite{SupernovaSearchTeam:1998fmf,SupernovaCosmologyProject:1998vns,Aghanim:2018eyx}, CMB (cosmic microwave bachground) fluctuations~\cite{Samtleben:2007zz,Planck:2018jri},weak lensing and
galaxy clustering~\cite{DES:2022ccp}
 and BAO (baryon acoustic oscillations) ~\cite{SDSS:2005xqv}, 
 indicate the accelerated expansion of the current universe.
 A simple explanation for the origin of the acceleration is given by the non-vanishing cosmological constant, $\Lambda_{\rm DE}$.
The late-time acceleration
is also explained by considering the potential energy of a scalar field, called quintessence~\cite{Fujii:1982ms,Zlatev:1998tr,Copeland:2006wr}.
Furthermore, Quintessential Inflation attempts to realize inflation and dark energy in a single scalar field of potential energy~\cite{Peebles:1998qn,deHaro:2021swo}.
However, it is difficult to explain the small energy scale of the cosmological constant in either way, i.e., the large hierarchy between the dark energy and the inflationary scale.
In this paper, we attempt to explore models that more naturally induce a quintessential inflation.

$F(R)$ modified gravity is a theory in which the Einstein-Hilbert action is modified into an arbitrary function of the Ricci scalar, $F(R)$. The theory
can be rewritten as an equivalent scalar-tensor theory through the conformal transformation~\cite{Nojiri:2006ri,Nojiri:2010wj,Nojiri:2017ncd,jordan1955schwerkraft}.
The Starobinsky model, which introduced an additional $R^2$ term, is one of the most famous models of the $F(R)$ modified gravity~\cite {Starobinsky:1979ty,Starobinsky:1980te}.
The early and late time accelerating expansions are caused by the scalar mode of gravity and the cosmological constant, respectively.
Palatini formalism and other $F(R)$ gravity theories address the dark energy explanation\cite{Oikonomou:2022wuk,Odintsov:2023cli,Montani:2023xpd} and quintessential inflation\cite{Dimopoulos:2020pas,Dimopoulos:2022rdp}.

Cartan $F(R)$ gravity is a natural extension of $F(R)$ gravity in Cartan formalism based on the Riemann-Cartan geometry formulated by the vierbein~\cite{cartan:1923}.
The important feature of Cartan $F(R)$ gravity is a non-vanishing torsion~\cite{Montesinos:2020pxv,Inagaki:2022blm}.
The curvature scalar $R$ is then divided into the part obtained from the Levi-Civita connection in General Relativity and a kinetic term for a scalar field obtained from the torsion.
Therefore, an equivalent scalar-tensor theory can be derived in Cartan $F(R)$ gravity.
It should be noted that metric-affine and Palatini $F(R)$ gravity are rewritten forms from a certain class of Brans-Dicke type scalar-tensor theories after conformal transformations~\cite{Capozziello:2007tj,Capozziello:2008yx,Sotiriou:2009xt,Capozziello:2009mq,Olmo:2011uz}.
On the other hand, the conformal transformation is not necessary to obtain an equivalent scalar-tensor theory in Cartan $F(R)$ gravity.
Thus Cartan $F(R)$ gravity is free from the ambiguity regarding the equivalence of physics before and after the conformal transformation~\cite{Catena:2006bd,Steinwachs:2011zs,Kamenshchik:2014waa,Hamada:2016onh}.

The purpose of this paper is to naturally realize a quintessential inflation scenario by using a logarithmic model in Cartan $F(R)$ gravity.
This paper is organized as follows.
In Sec.~\ref{section::Cartan gravity} we briefly introduce the Cartan formalism and Cartan $F(R)$ gravity.
It is shown that the logarithmic model has a potential for accelerating expansion of the Universe on two energy scales.
In Sec.~\ref{section:quintessential inflation} quintessential inflation scenario is considered.
First, we employ the standard slow-roll scenario and calculate the CMB fluctuations in the logarithmic model of Cartan $F(R)$ gravity.
Then, the reheating process was calculated analytically and numerically.
Next, the quintessence scenario for dark energy is adapted in the logarithmic model.
In Sec.~\ref{section::conculusion} we will give some concluding remarks.

\section{Cartan $F(R)$ gravity}
\label{section::Cartan gravity}
\subsection{Cartan $F(R)$ gravity}

There are several variations in the geometric formulation of gravity.
The theory of gravity is reformulated
on the Riemann-Cartan geometry described by the vierbein ${e^i}_\mu$ and the spin connection ${\omega^{ij}}_\nu$.
The vierbein connects the curved metric $g_{\mu\nu}$ and flat one $\eta_{ij}$ with,
\begin{align}\label{Eq:Definition of tetrad}
g_{\mu\nu}=\eta_{ij}{e^i}_\mu{e^j}_\nu.
\end{align}
The action of Cartan $F(R)$ gravity is defined by replacing the curvature scalar $R$ 
in Einstein-Cartan theory with a general function $F(R)$,
\begin{align}\label{eqs:action FR}
S=\int d^4xe\left( \frac{{M_{\rm Pl}}^2}{2}F(R)+\mathcal{L}_{\rm{m}}\right),
\end{align}
where $M_{\rm Pl}$ indicates the Planck scale and a volume element is given by the determinant of the vierbein, $e$.
The curvature scalar is expressed by the spin connection and the vierbein,
\begin{align*}
R
= {e_i}^\mu{e_j}^\nu {R^{ij}}_{\mu\nu}(\omega,\partial\omega)
= {e_i}^\mu{e_j}^\nu\left[\partial_{\mu}{\omega^{ij}}_\nu-\partial_{\nu}{\omega^{ij}}_\mu+{\omega^i}_{k\mu}{\omega^{kj}}_\nu-{\omega^i}_{k\nu}{\omega^{kj}}_\mu\right].
\end{align*}

In Riemann-Cartan geometry, a geometric tensor, ${T^{\rho}}_{\mu\nu}$, called torsion is introduced,
\begin{align*}
{T^{\rho}}_{\mu\nu}\equiv{\Gamma^\rho}_{\mu\nu}-{\Gamma^\rho}_{\nu\mu},
\end{align*}
where Affine connection is given by ${\Gamma^\rho}_{\mu\nu}= {e_a}^\rho D_\nu {e^a}_\mu$ and $D_\nu$ is the covariant derivative for the local Lorentz transformation,
\begin{align*}
D_\nu {e^k}_\mu=\partial_\nu {e^a}_\mu+{{\omega^k}_{l\nu}}{e^l}_\mu.
\end{align*}
The Affine connection is not necessarily invariant under the replacement of the lower indices, ${\Gamma^\rho}_{\mu\nu}\neq{\Gamma^\rho}_{\nu\mu}$.
The torsion is represented by the derivative of $F(R)$ and the vierbein from the Cartan equation \cite{Montesinos:2020pxv,Inagaki:2022blm};
\begin{align}\label{eqs:torsion of FR}
{T^k}_{ij}=\frac{1}{2}({\delta^k}_j{e_i}^\lambda-{\delta^k}_i{e_j}^\lambda)\partial_\lambda\ln F'(R),
\end{align}
where we assume that the matter field is independent of the spin connection.
It should be noted that the torsion vanishes in Einstein-Cartan theory, $F(R)=R$.
Non-vanishing torsion can be obtained in Cartan $F(R)$ gravity.
Torsion can be extracted from the curvature scalar,
\begin{align}\label{eqs:Ricci scalar into non and torsion2}
R=R_E+T-2{\nabla_E}_\mu T^\mu,
\end{align}
where the subscript $E$ in $R_E$ and $\nabla_E$ stands for the Ricci scalar and the covariant derivative given by the Levi-Civita connection.
$T_\mu$ represents the torsion vector $T_\mu={T^\lambda}_{\mu\lambda}$ and the torsion scalar $T$ is defined to contract the torsion and torsion vector as
\begin{align*}
T=\frac{1}{4}T^{\rho\mu\nu}T_{\rho\mu\nu}-\frac{1}{4}T^{\rho\mu\nu}T_{\mu\nu\rho}-\frac{1}{4}T^{\rho\mu\nu}T_{\nu\rho\mu}-T^\mu T_\mu.
\end{align*}
Thus, the curvature scalar is divided into the non-torsion part, i.e. Ricci scalar in general relativity, $R_E$, and an additional torsion part. 
Substituting Eq.(\ref{eqs:torsion of FR}) into Eq.(\ref{eqs:Ricci scalar into non and torsion2}),
the curvature scalar is rewritten as
\begin{align}\label{eqs:Ricci scalar to einstein and torsion}
R=R_E-\frac{3}{2}\partial_{\lambda}\ln F'(R)\partial^{\lambda}\ln F'(R)-3 \nabla_E^2 \ln F'(R).
\end{align}

Now we consider a certain class of Cartan $F(R)$ gravity expressed as $F(R)=R+f(R)$,
\begin{align}
S =\int d^4x e\frac{{M_{\rm Pl}}^2}{2}\left(R+f(R)\right) .
\label{eqs:action for f}
\end{align}
We define the scalaron field, $\phi$, as
\begin{align}\label{eqs:def scalar field}
\phi\equiv-\sqrt{\frac{3}{2}}{M_{\rm Pl}}\ln F'(R).
\end{align}
and insert it into Eq.(\ref{eqs:action for f}) with Eq.(\ref{eqs:Ricci scalar to einstein and torsion}), 
the gravity part of the action is rewritten
to be the Einstein-Hilbert term and the scalaron field,
\begin{align}
S 
= \int d^4x e\left(\frac{{M_{\rm Pl}}^2}{2}R_E-\frac{1}{2}\partial_{\lambda}\phi \partial^{\lambda}\phi-V(\phi)\right).
\label{eqs:action for R and scalar}
\end{align}
We assume that $\ln F'(R)$ vanishes at a distance,
consequently, a total derivative, the last term in \eqref{eqs:Ricci scalar to einstein and torsion}, is omitted.
Then the potential, $V(\phi)$, is given by,
\begin{align} \label{eqs:potential for phi in general}
V(\phi) \equiv -\frac{{M_{\rm Pl}}^2}{2}\left. f(R)\right|_{R=R(\phi)}.
\end{align}
The potential is represented as a function of the scalaron field $\phi$ with Eq.(\ref{eqs:def scalar field}).

Thus, the equivalent scalar-tensor theory (\ref{eqs:action for R and scalar}) is derived without any conformal transformations.
The potential, $V(\phi)$, has a simpler representation from the one in the scalar-tensor theory obtained from conventional $F(R)$ gravity after the conformal transformation~\cite{Nojiri:2017ncd}.
Various potentials can be obtained from the function $f(R)$ in Cartan $F(R)$ gravity.
As an example, $f(R)=-R^2$ in Cartan $F(R)$ gravity derives a potential equal to the Starobinky model~\cite{Inagaki:2022blm}.

\subsection{Logarithmic model}
\label{section::Log model}
\begin{figure}
    \centering
    \includegraphics[width=0.7\linewidth]{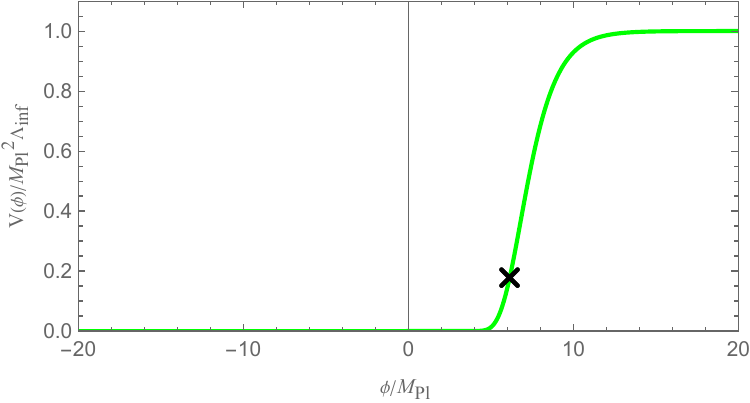}
    \caption{The Green line is the potentials of the logarithmic model \eqref{Model::Logarithmic} with $\alpha=0.003875$.
    Cross marks($\times$) show the end of inflation, the point at which the slow-roll parameter becomes one, $\epsilon_V=1$.}
   \label{Model::Log::Potential}
\end{figure}
In Ref.~\cite{Inagaki:2023dzn} various models of Cartan $F(R)$ gravity have been investigated. It has been found that the potential of the logarithmic model has a long tail after the end of the inflation era.
In this paper, we focus on the quintessential inflation in the logarithmic model. 

The logarithmic model is defined by
\begin{align}
    f(R) = -\alpha R \ln \Big( 1 + \frac{R}{R_0} \Big),
    \label{Model::Logarithmic}
\end{align}
with two parameters, $\alpha$ and $R_0$.
This model is inspired by the context of quantum field theory (QFT) that mimics logarithmic one-loop corrections.
At the weak curvature limit, the logarithmic model deforms the corrections to keep the Einstein-Hilbert action.
Several variations of the logarithmic corrections have been investigated in Ref.~\cite{Nojiri:2003ni}.

As the coupling $\alpha$ increases, the potential obtained from Eq.~\eqref{Model::Logarithmic} approaches the Starobinsky model, $f(R)=-R^2$~\cite{Inagaki:2023dzn}.
\begin{figure}
     \centering
     \includegraphics[width=0.7\linewidth]{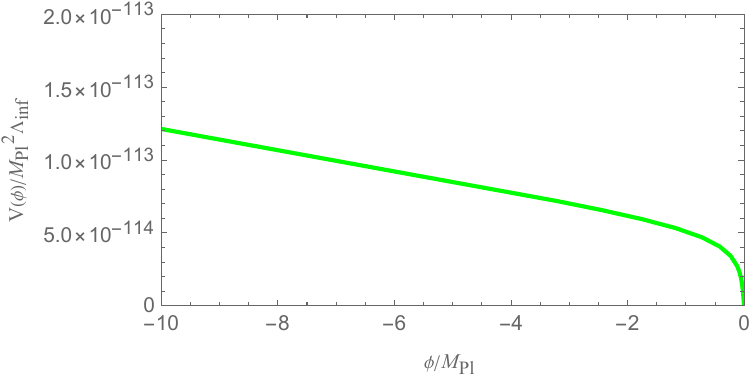}
     \caption{The Green line is the potential with $\alpha=0.003875$ at $\phi<0$. }
     \label{Model::Logarithmic::Potential::negative}
\end{figure}
For a small coupling, the model is approximated as $f(R)\sim R^{1+\alpha}$ for a positive $\phi$ and 
the potential is given by
\begin{align}\label{eqs:potential_positive}
V(\phi)\propto\left(1-e^{-\sqrt{\frac{2}{3}}\frac{\phi}{{M_ {\rm Pl}}}}\right)^{1+\frac{1}{\alpha}}.
\end{align}

In Fig.~\ref{Model::Log::Potential},
we show the potential of the logarithmic model.
It can be shown that the potential has a flat plateau for a large $\phi$. We assume that the potential~(\ref{eqs:potential for phi in general}) has the inflationary scale, $\Lambda_{\rm Inf}$ at the plateau and adopts the large-field inflation scenario.
From Eq.~\eqref{eqs:def scalar field} the function, $F(R)$, satisfies
\begin{align}
    F'(R)\sim 0,
\end{align}
at the plateau.
Thus, we obtain the relationship between the model parameters, $R_0$ and $\alpha$,
\begin{align}\label{eqs::model parameter}
    R_0= 2\Lambda_{\rm Inf} e^{-\frac{1}{\alpha}}.
\end{align}
The small coupling, $\alpha$, makes the parameter, $R_0$, exponentially smaller than the inflationary scale.

To derive the approximate potential for a negative $\phi$ we adapt,
\begin{align}
    \frac{R}{R_0}=-1+e^{-\zeta}\ (0\leq\zeta<\infty),
\end{align}
to the expression for the scalaron field (\ref{eqs:def scalar field}) and obtain
\begin{align*}
\phi/{M_{\rm Pl}}&=-\sqrt{\frac{3}{2}}\ln\left(1-\alpha+\alpha e^{\zeta}+\alpha\zeta \right)
\\
&\rightarrow -\sqrt{\frac{3}{2}}\zeta,
\hspace{20mm}\mbox{for a large $\zeta$}.
\end{align*}
Under this approximation, the potential is expressed as
\begin{align} 
V(\phi)\sim \frac{\alpha R_0}{2}\zeta 
= \sqrt{\frac{2}{3}}\alpha\Lambda_{\rm Inf}e^{-\frac{1}{\alpha}}\left(\frac{-\phi}{{M_{\rm Pl}}}\right),
\label{pot:linear}
\end{align}
where we use the relationship (\ref{eqs::model parameter}).
Therefore, the potential energy is exponentially suppressed for a small $\alpha$ in the negative $\phi$ region.
The dark energy scale is obtained for $\alpha\sim0.0039$,
\begin{align}
    \sqrt{\frac{2}{3}}\alpha\Lambda_{\rm Inf}e^{-\frac{1}{\alpha}}
    \sim\frac{\Lambda_{\rm Inf}}{10^{114}}\sim \Lambda_{\rm DE},
\end{align}
where the inflationary scale is taken to be $\Lambda_{\rm Inf}\sim10^{15}{\rm GeV}$.
At this time, the potential of the logarithmic model 
introduces
the energy scale gap between inflation and dark energy, $\Lambda_{\rm Inf}/\Lambda_{\rm DE}\sim 10^{114}$.

The potential (\ref{eqs:potential for phi in general}) for the negative $\phi$ region is numerically calculated and shown in Fig.~\ref{Model::Logarithmic::Potential::negative}.
We observe approximately linear behavior for a small $\phi$ which is consistent with Eq.~\eqref{pot:linear}.
As we have seen, the logarithmic model with a small coupling realizes both inflationary and dark energy scale potential energy.
Below we study specifically the inflation, reheating, and dark energy processes.

\section{Quintessential Inflation}
\label{section:quintessential inflation}
In this section, we evaluate the evolution of the universe in the logarithmic model of Cartan $F(R)$ gravity.
In the previous section, we have shown that the potential of the model develops the inflationary and dark energy scale in the positive and the negative $\phi$ region, respectively.
Thus, the model has the potential to explain the accelerated expansion of the universe in two eras.
The attempt to derive the accelerated expansion in inflation and dark energy eras is called Quintessential Inflation~\cite{Peebles:1998qn}.
We employ the flat FLRW (Friedmann‐Lemaître‐Robertson-Walker) metric and 
study the phenomena in the inflation, reheating, and dark energy eras caused by the scalaron field induced by Cartan $F(R)$ gravity.

\subsection{Inflation}
\label{section::inflation}
\begin{figure}
\centering
      \includegraphics[width=0.7\linewidth]{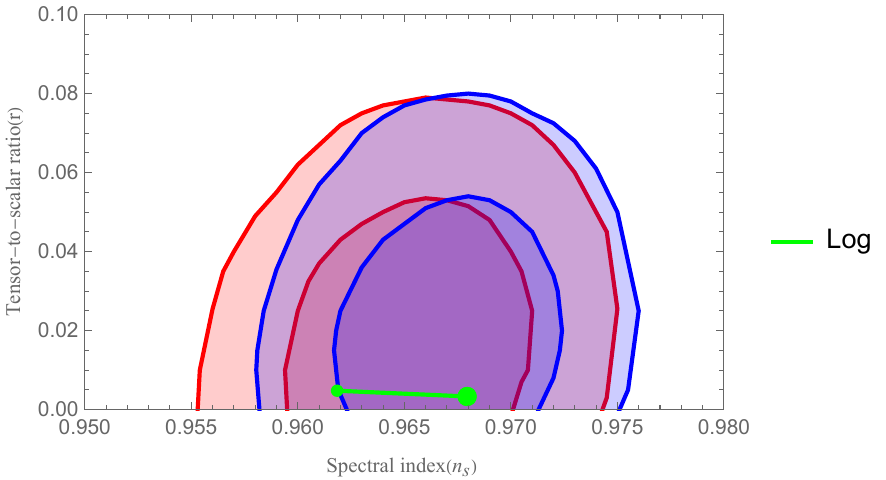}
    \caption{Numerical result: The green line shows the  CMB fluctuations for the logarithmic models with $\alpha=0.003875$ at $N=50$ (Left) and $N=60$ (Right) with Planck constraints. The blue and red areas indicate the Planck 2018 constraints in  \cite{Planck:2018jri}.}
       \label{Fig::CMBflu::Plancl}
\end{figure}
The slow-roll inflation is a standard scenario of the early-time accelerating expansion of the Universe.
Here, we adapt the slow-roll scenario of inflation to the logarithmic model of Cartan $F(R)$ gravity.
We identify the scalaron field, $\phi$, as the inflaton and assume that the scalaron potential energy dominates the energy density of the Universe.

In the slow-roll scenario, the inflation is controlled by the slow-roll parameters, $\epsilon_V, \eta_V$. 
These parameters are  described as a function of $R$ through the form of the potential,
\begin{align}
    \label{epsilonv::slow-roll}
    \varepsilon_V&=\frac{{M_{\rm Pl}}^2}{2}\left(\frac{V'}{V}\right)^2
   = {
        \frac{1}{3} \Big(\frac{F'}{F''}\Big)^2
        \Big(\frac{f'}{f}\Big)^2
    },
    \\
    \label{etav::slow-roll}
    \eta_V
    &=\frac{{M_{\rm Pl}}^2 V''}{V}
    = \frac{2}{3} \Big(\frac{F'}{F''}\Big)^2
    \frac{f'}{f} \Big\{
        \frac{f''}{f'} + \frac{F''}{F'}
        - \frac{F'''}{F''}
    \Big\},
\end{align}
Inflation lasts while the slow-roll parameters are less than 1 and ends when they approach unity, in the current model, $\varepsilon_V\sim 1$.
The e-folding number $N$ during the inflation is given by 
\begin{align}
    N
    = \frac{3}{2} \int^{R_*}_{R_\text{end}} dR \Big(\frac{F''}{F'}\Big)^2
    \frac{f}{f'},
    \label{eFoldingNumber::generalForm}
\end{align}
where $R_*$ and $R_\text{end}$ denote the curvature at the beginning and end of inflation.
The latter value is obtained from the condition, $\varepsilon_V = 1$. 
The former one is evaluated to obtain a suitable e-folding number $N = 50\sim 60$,
which is required to solve the horizon and flatness problems.

The quantum fluctuations of the inflaton are converted to initial values of the curvature perturbations by accelerating expansion.
The classical fluctuations affect CMB fluctuations characterized by the power spectrum, $A_s$, the spectral index $n_s$, and the scalar-tensor ratio $r$. The inflationary parameters are estimated by
\begin{align*}
  A_s
   &=
        -\frac{f}{16 \pi^2 M_\text{Pl}^2} \Big(\frac{F''}{F'}\Big)^2
        \Big(\frac{f}{f'}\Big)^2,
    \\
    n_s
        &= 1 - 6\varepsilon_V + 2 \eta_V,
    \\ 
    r &=16\varepsilon_V.
\end{align*}
These predictions
should satisfy the constraints of Planck 2018~\cite{Planck:2018jri}.
From the power spectrum, the inflationary scale is fixed to be $\Lambda_{\rm Inf}\sim10^{15}{\rm GeV}$. 
The spectral index and scalar-tensor ratio constrain the models of inflation.
The detailed analysis for Cartan $F(R)$ gravity has been discussed in Ref.\cite{Inagaki:2023dzn}.
It has been found that the logarithmic model has robustness about the model parameter variations.
Figure~\ref{Fig::CMBflu::Plancl} shows the numerical results on CMB fluctuations.
It is observed that the model satisfies the constraints of Planck 2018. 
Thus, we conclude that the logarithmic model in Cartan $F(R)$ gravity is a candidate of the model of inflation.

\subsection{Reheating}
\label{section::reheating}
During inflation, the adiabatic expansion of the universe rapidly lowers the temperature to almost zero.
The universe should be reheated after the inflation to realize the process 
such as nucleosynthesis.
In Cartan $F(R)$ gravity the reheating process is induced by converting the energy of the scalaron field into radiation. 
We assume an interaction between the scalaron field and radiation characterized by $\Gamma$.
The interaction acts as a frictional force term in the equation of motion for the scalaron field,
\begin{align}\label{eqs::scalar::eom}
\ddot{\phi}+(3H+\Gamma)\dot{\phi}+V'(\phi)=0,
\end{align}
where $H$ is Hubble parameter defined by $H=\dot{a}(t)/a(t)$ with the scale factor of the universe, $a(t)$. 
It is evaluated by Friedmann equation,
\begin{align}\label{eqs::Friedmanneq}
3H^2M_{\rm Pl}^2=\rho_\phi+\rho_r,
\end{align}
where $\rho_\phi$ and $\rho_r$ are the energy density of the scalaron field and radiation, respectively. The energy density, $\rho_\phi$, is a total of the scalaron kinetic energy and the potential energy,
\begin{align}
\rho_\phi=\frac{1}{2}\dot{\phi}^2+V(\phi).
\end{align}
The energy density of radiation, $\rho_r$, evolves according to the differential equation,
\begin{align}\label{eqs::scalar::radiation}
\dot{\rho}_r+4H\rho_r=\Gamma\rho_\phi.
\end{align}

As shown in Fig.\ref{Model::Log::Potential}, the potential of the logarithmic model has two widely separated energy scales.
The region, $0<\phi<5$, looks like the gentle downhill with the inflation scale, and the negative region, $\phi<0$, is the linear slope with the dark energy scale.
The scalaron field does not oscillate at the bottom of the potential. 

In conventional cases such as the $R^2$ model, the average over time of the kinetic and potential energy is equal due to the virial theorem.
After the end of the inflation in the logarithmic model, the potential energy decreases very small and the kinetic energy comes to dominate the universe.
This situation is called kination and is described as
\begin{align*}
\rho_\phi\sim\frac{1}{2}\dot{\phi}^2.
\end{align*}
Neglecting the potential term in Eq.~\eqref{eqs::scalar::eom}, we derive the equation for the energy density, $\rho_\phi$,
\begin{align*}
\dot{\rho}_\phi+(6H+2\Gamma)\rho_\phi=0.
\end{align*}
The solution is given by
\begin{align}
\rho_\phi(t)=\rho_\phi^{(i)}\left(\frac{a(t)}{a^{(i)}}\right)^{-6}e^{-2\Gamma(t-t_i)}.
\end{align}
where $t_i$ is the start time of the reheating process and we write $\rho_\phi^{(i)}\equiv \rho_\phi(t_i)$ and $a^{(i)}\equiv a(t_i)$.
For a small $\Gamma$ the energy density, $\rho_\phi$, is proportional to $a(t)^{-6}$ and it dominates the right hand side of Friedmann equation~\eqref{eqs::Friedmanneq}. Since the scale factor $a(t)^{-6}$ is proportional to $t^{-2}$ from the Friedmann equation, we write
\begin{align*}
\frac{a(t)}{a^{(i)}}=\left(\frac{t}{t_i}\right)^{1/3}.
\end{align*}
Then Eq.\eqref{eqs::scalar::radiation} is rewritten as
\begin{align}
\dot{\rho}_r +4H\rho_r =\Gamma \rho_{\phi}^{(i)}\left(\frac{t}{t_i}\right)^{-2}.
\end{align}
This differential equation can be solved analytically. The solution is found to be
\begin{align}\label{eqs::radiationdensity}
\rho_r(t)=\frac{3\Gamma\rho_{\phi}^{(i)} t_i^2}{t}\left[1-\left(\frac{t}{t_i}\right)^{-\frac{1}{3}}\right].
\end{align}
From Eq.~\eqref{eqs::radiationdensity}, the maximum value of the radiation energy density is given by $\rho_r=(81/256)\Gamma\rho_{\phi}^{(i)} t_i\simeq 0.316\Gamma\rho_{\phi}^{(i)} t_i$ at the time, $t=(64/27)t_i\sim 2.37t_i$.
After the maximum value, the radiation energy density decreases with $\rho_r\propto 1/t$ as the universe expands.

We estimate the reheating time, $t_R$, by solving $\rho_r(t_R)=\rho_\phi(t_R)$. Ignoring the second term in the parenthesis in Eq.~\eqref{eqs::radiationdensity}, we obtain
\begin{align}\label{eqs::reheat time}
  t_R=\frac{1}{3\Gamma}. 
\end{align}
The radiation density at $t_R$ is given by
\begin{align}\label{rhor:tr}
\rho_r(t_R)&=9\Gamma^2 \rho_{\phi}^{(i)} t_i^2
=27\Gamma^2{M_{\rm Pl}}^2,
\end{align}
where we set $\rho_{\phi}^{(i)}=3 H^2(t=t_i)M^2_{\rm Pl}$ and $t_i = 1/H(t=t_i)$.
According to Stefan-Boltzmann law, the radiation energy density is represented as a function of temperature,
\begin{align}\label{rhor:SB}
\rho_r=\frac{\pi^2g_*(k_{\rm B}T)^4}{30}.
\end{align}
Comparing Eqs.~\eqref{rhor:tr} and \eqref{rhor:SB}, we estimate the reheating temperature, $T_R$, for the kination,
\begin{align}\nonumber
k_{\rm B}T_R&=\left(\frac{30\cdot27}{\pi^2g_*}\right)^{\frac{1}{4}}\sqrt{{M_{\rm Pl}}\Gamma}
\\ \label{eqs::reheat temp}
&\simeq 1.56\times10^9 \left(\frac{106.75}{g_*}\right)^{\frac{1}{4}}\sqrt{\frac{\Gamma}{1{\rm GeV}}}\ [{\rm GeV}].
\end{align}
\begin{figure}
    \begin{minipage}{0.48\linewidth}
    \centering 
    \includegraphics[,scale=0.48]{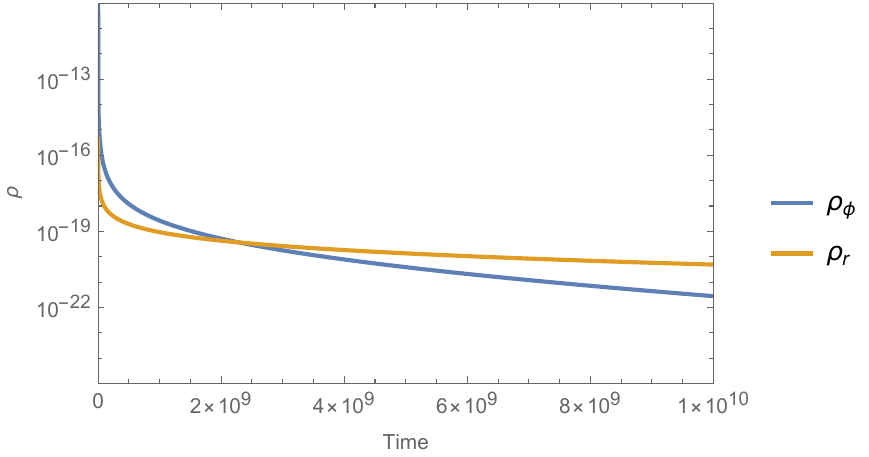}
    \subcaption{Time dependence of density parameter.}
    \end{minipage}
    \begin{minipage}{0.48\linewidth}
    \centering
    \includegraphics[scale=0.48]{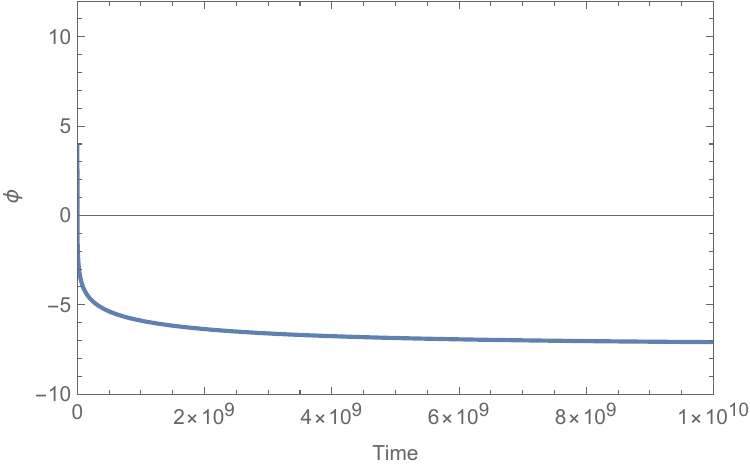}
    \subcaption{Time dependence of the scalaron field.}
    \end{minipage}
    \caption{Numerical results of the reheating process in the logarithmic model with $\alpha=0.003875$. 
       Time is scaled by $\tilde{t}={\Lambda_{\rm Inf}}^{\frac{1}{2}}t$ and $\Gamma/{\Lambda_{\rm Inf}}^{\frac{1}{2}}=10^{-10}$. 
       The density is scaled by $\rho/M_{\rm Pl}^2\Lambda_{\rm Inf}$.(a)The Blue lines represent the energy density of the scalaron field. The orange line represents the energy density of radiation.
       (b)The blue line shows the time dependence of the scalaron field with the initial value, $\phi_0/M_{\rm Pl}=11.3$.} 
    \label{Fig::Reheating}
\end{figure}

We numerically solve the differential equations for the energy density and show the result in Fig.\ref{Fig::Reheating}.
In this calculation, we suppose $\Gamma/{\Lambda_{\rm Inf}}^{1/2}=10^{-10}$, and set the reheating start time at $\Lambda_{\rm Inf} t_i=10^3$ to fit the Hubble parameter, $H(t=t_i)\sim 10^{12}{\rm GeV}$.
\begin{table}
\begin{center}
\begin{tabular}{ccc}
    \hline
    $\Gamma/\Lambda_{\rm Inf}$ & $\Lambda_{\rm Inf} t_R$ & $k_BT_R$ [GeV]
    \\ \hline
    $10^{-10}$ & $2.29\times10^9$ & $2.78\times10^{11}$
    \\ \hline
\end{tabular}
\caption{Reheating time and temperature}\label{table:tR}
\end{center}
\end{table}
The numerical results (Tab.~\ref{table:tR})
almost coincide with the analytic calculation in Eqs.\eqref{eqs::reheat time} and \eqref{eqs::reheat temp}.
The reheating temperature is sufficiently higher than the one of the Big-Bang nucleosynthesis, $T_{\rm BBN}\sim 1{\rm MeV}$. 
 Therefore, we conclude that the logarithmic model in Cartan $F(R)$ gravity implements the reheating process.

\subsection{Dark energy}
\label{section::dark energy}
After the reheating process, the universe proceeds to the hot Big Bang scenario.
The leading role in the evolution of the universe shifts from the scalaron field to radiation and matter.
In the radiation- or matter-dominant era, the scalaron field departs from its role in expanding the universe.
The radiation or matter mainly contributes to the Hubble parameter through Friedmann equation~\eqref{eqs::Friedmanneq}. 
Since the second term, $(3H+\Gamma)\dot{\phi}$, in Eq.~\eqref{eqs::scalar::eom} acts as a friction term, the scalaron field speeds down and "freezes" at the slope.
On the other hand, the energy density of radiation and matter decreases with the expansion of the universe.
Here, we construct the quintessence scenario in which the scalaron field dominates the universe again on the dark energy scale. 

For practical calculations, it is convenient to introduce the density parameter $x$, $y$ for the kinetic and potential energy of the scalaron field: 
\begin{align}\label{def:x velocity of phi}
x\equiv \frac{\dot{\phi}}{\sqrt{6}{M_{\rm Pl}}H},
\\ \label{def:y potential}
y\equiv \frac{\sqrt{V}}{\sqrt{3}{M_{\rm Pl}}H},
\end{align}
and the density parameter of radiation, matter, and scalaron:
\begin{align*}
\Omega_{I}\equiv \frac{\rho_I}{3{M_{\rm Pl}}^2H^2}\ (I={r},{m},\phi).
\end{align*}
The density parameter of the scalaron field, $\Omega_{\phi}$, is the sum of the kinetic and potential parts,
\begin{align*}
\Omega_{\phi}=x^2+y^2.
\end{align*}
Friedmann equation~\eqref{eqs::Friedmanneq} constrains the sum of the density parameters,
\begin{align}\label{eqs::Fridmann::densitypara}
1&=x^2+y^2+\Omega_r+\Omega_m.
\end{align}
The time derivative of ~\eqref{eqs::Friedmanneq} derives 
\begin{align}\label{eqs::1sthubblepara}
 \frac{\dot{H}}{H^2}
=-\frac{1}{2}\left(3+3x^2-3y^2+\Omega_r\right)
\end{align}
where $\dot{H}$ represents the time derivative of $H$.

We can use the redshift, $z$ instead of time.
From the relationship between the redshift and the scale factor, $a=1/(z+1)$, we obtain
 \begin{align*}
 &\ln \frac{a}{a_0}=-\ln (z+1),
 \end{align*}
where $a_0$ is the scale factor of the current Universe and the current redshift  
is $z=0$.
The time derivative is translated to the redshift,
\begin{align*}
\dot{f}=\frac{d\ln a}{dt}\frac{d}{d\ln a}f
=-H\frac{d}{d\ln (z+1)}f.
\end{align*}
To derive the redshift dependence of the density parameters, we start with the following  equations,
\begin{align}\label{eqs::scalar::denpara}
    &\dot{\rho_\phi}+3H\dot{\phi}^2=0,
    \\ \label{eqs::radiation::denpara}
    &\dot{\rho_r}+4H\rho_r=0,
    \\ \label{eqs::matter::denpara}
    &\dot{\rho_m}+3H\rho_m=0.
\end{align}
Using Eq.\eqref{eqs::scalar::eom} with $\Gamma = 0$ and Eqs.~(\ref{def:x velocity of phi})
-\eqref{eqs::matter::denpara}, we derive the equations on $x, y$ and the density parameter for radiation, $\Omega_r$,
\begin{align}\label{eqs::x}
 \frac{d\phi}{d\ln (z+1)}&=-\sqrt{6}M_{\rm Pl}x,
 \\
\frac{dx}{d\ln (z+1)}&=-\frac{x}{2}(3x^2-3y^2-3+\Omega_r)-\frac{\sqrt{6}}{2}\lambda y^2,
\\
\frac{dy}{d\ln (z+1)}&=-\frac{y}{2}(3x^2-3y^2+3+\Omega_r-\sqrt{6}\lambda x),
\\
\frac{d\Omega_r}{d\ln (z+1)}
&=\Omega_r(\Omega_r-1+3x^2-3y^2),
\end{align}
with
\begin{align}
\lambda=-\frac{{M_{\rm Pl}}V'}{V}.
\end{align}
The evolution of the universe is determined by solving these equations under the constraint Eq.~\eqref{eqs::Fridmann::densitypara}.
The equation of State (EoS) parameter, $\omega_{\rm{eff}}$, is also evaluated to determine whether the quintessence causes of the late-time expansion,
\begin{align}
\omega_{\rm{eff}}=\frac{x^2-y^2+\frac{1}{3}\Omega_r}{x^2+y^2+\Omega_r+\Omega_m}.
\end{align}
The EoS parameter for the scalaron field is defined by
\begin{align}
\omega_{0}=\frac{x^2-y^2}{x^2+y^2}.
\end{align}
This parameter indicates the scalaron field contribution.

\begin{figure}
    \begin{minipage}{1\linewidth}
    \centering 
    \includegraphics[,scale=0.8]{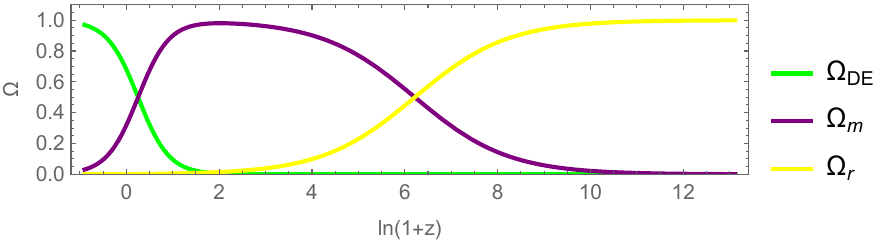}
    \subcaption{Redshift evolution of density parameter.}
    \end{minipage}
    \\
    \begin{minipage}{1\linewidth}
    \centering
    \includegraphics[scale=0.8]{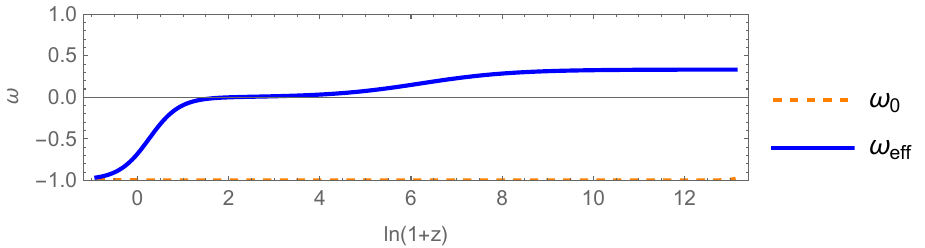}
    \subcaption{Redshift evolution of the scalaron field.}
    \end{minipage}
    \caption{Redshift($z$) evolution of the density parameter. 
    The blue line shows the EoS parameter. The dashed orange line shows the EoS parameter for the scalaron field. The initial condition is $\Omega_r= 0.999, x=1.31\times10^{-11}, 
y= 1.31\times10^{-10}, \phi_0= -7 M_{\rm Pl}$ at $z=5\times 10^5$.} 
    \label{Fig::Dark energy}
\end{figure}

We numerically calculate the density and EoS parameters and plot the results in Fig.~\ref{Fig::Dark energy}.
The initial conditions are determined to fit the current observations, $\Omega_{\rm DE}(z=0)=0.68, \Omega_{\rm m}(z=0)=0.32$~\cite{Planck:2018jri}.
From the results of the reheating era, we choose the initial conditions, $\Omega_r= 0.999, x=1.31\times10^{-11}, 
y= 1.31\times10^{-10}$ at $z=5\times10^5$ as typical values in the logarithmic model of Cartan $F(R)$ gravity.
\begin{figure}
    \centering
    \includegraphics[scale=0.7]{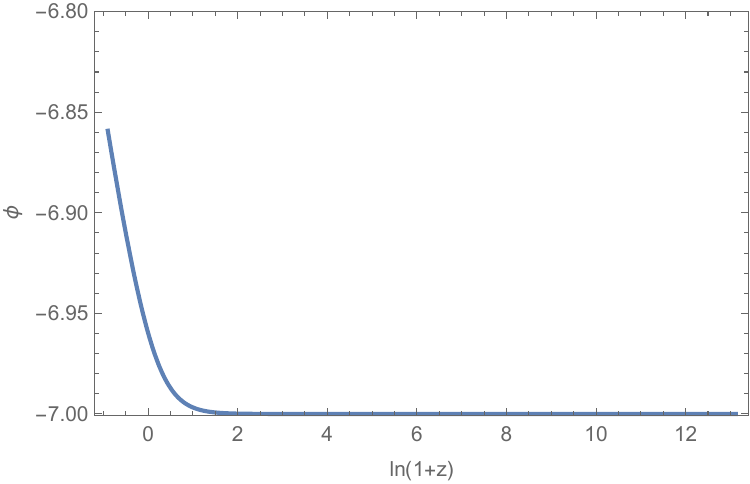}
    \caption{Redshift evolution of a scalaron field. The blue line shows the value of the scalaron field in terms of redshift. The initial value of the scalaron field is $\phi_0=-7 M_{\rm Pl}$.}
    \label{fig:redshift_phi}
\end{figure}
From the behavior of the density parameters, it can be seen that the scalaron field again dominates the energy density of the universe, following the radiation and matter-dominant eras.
The EoS parameter, $\omega_{\rm{eff}}$ gradually decreases and the contribution of the scalaron field is approximately constant.
The current EoS parameter is found to be $\omega_0(z=0)=-0.997$ $(\omega_{\rm eff}(z=0)=-0.679)$.
It is consistent with the current observational constraints $\omega_0<-0.95$~\cite{Planck:2018jri} and around $\omega_0\sim-1$~\cite{DES:2022ccp}.

We also evaluate the behavior of the scalaron field in the radiation and matter dominant era starting from the final value in the reheating process, $\phi_0=-7 M_{\rm Pl}$, in Fig.~\ref{Fig::Reheating}.
The numerical result is shown in Fig.~\ref{fig:redshift_phi}.
Since the scalaron field variable is almost fixed at $\phi=-7 M_{\rm Pl}$ for $\ln (1+z)>2$, we confirm that the scalaron field freezes in radiation and matter dominant era and then restarts the slow-roll process. Therefore, the current expansion is explained in analogy with quintessence in Cartan $F(R)$ gravity.

The numerical analysis predicts the future of the universe.
The scalaron field descends down the potential toward $\phi=0$.
We adopt the slow roll scenario to the current accelerated expansion and obtain
\begin{align}
\frac{d\phi}{d\ln (z+1)}\sim M_{\rm Pl}\frac{V'}{V}
=\frac{M_{\rm Pl}}{\phi}.
\end{align}
Solving this equation, we estimate the redshift at the end of the current accelerated expansion as $\phi=0$,
\begin{align}
    z=-1+e^{-\frac{{\phi_0}^2}{2M_{\rm Pl}^2}},
\end{align}
where $\phi_0$ indicates the current value of the scalaron field at $z=0$. Substituting the typical value, $\phi_0=7M_{\rm Pl}$, we get $z+1\sim2\times10^{-11}$. 
At that time, the singularity of space-time will not appear.

It should be noted the case in which the decay width, $\Gamma$, is large.
A large decay width introduces a strong friction in the motion of the scalaron field after the end of the inflation. The scalaron field stays at a positive value where the potential is given by Eq.~\eqref{eqs:potential_positive}.
\begin{figure}
    \begin{minipage}{0.48\linewidth}
    \centering 
    \includegraphics[,scale=0.48]{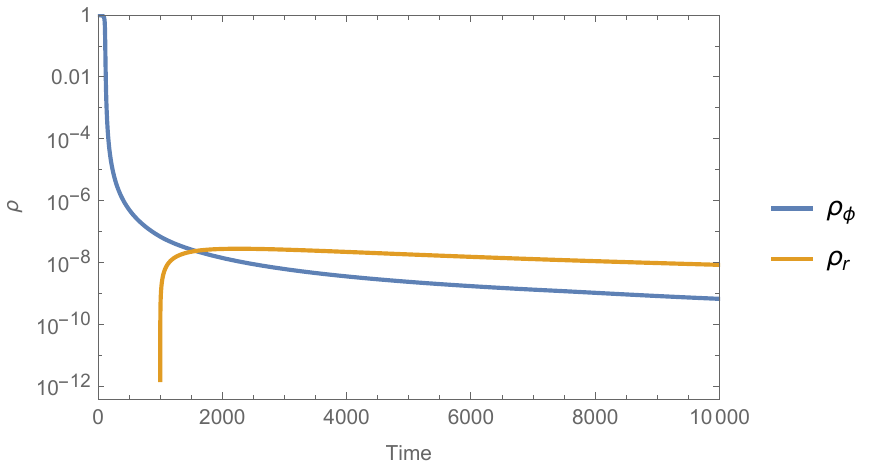}
    \subcaption{Time dependence of density parameter.}
    \end{minipage}
    \begin{minipage}{0.48\linewidth}
    \centering
    \includegraphics[scale=0.48]{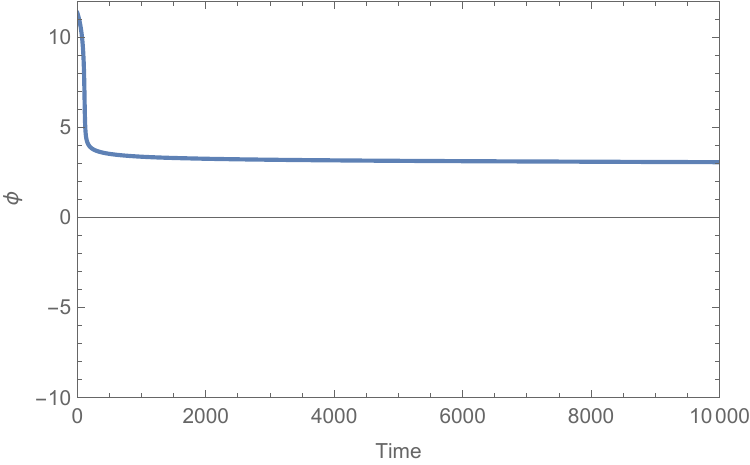}
    \subcaption{Time dependence of the scalaron field.}
    \end{minipage}
    \caption{Numerical results of the reheating in the logarithmic model with $\alpha=0.003875$. 
       Time is scaled by $\tilde{t}={\Lambda_{\rm Inf}}^{\frac{1}{2}}t$ and $\Gamma/{\Lambda_{\rm Inf}}^{\frac{1}{2}}=10^{-3}$.
       The density is scaled by $\rho/M_{\rm Pl}^2\Lambda_{\rm Inf}$.(a)The Blue lines represent the energy density of the scalaron field. The Orange line represents the energy density of radiation.
       (b)The blue line shows the time dependence of the scalaron field with the initial value, $\phi_0/M_{\rm Pl}=11.3$.} 
    \label{Fig::Reheating_large}
\end{figure}
\begin{figure}
    \begin{minipage}{1\linewidth}
    \centering 
    \includegraphics[,scale=0.8]{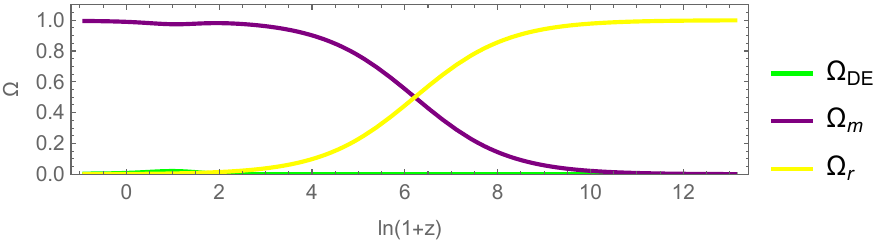}
    \subcaption{Redshift evolution of density parameter.}
    \end{minipage}
    \\
    \begin{minipage}{1\linewidth}
    \centering
    \includegraphics[scale=0.8]{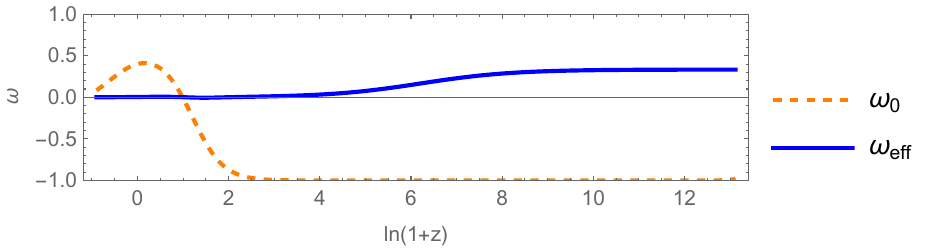}
    \subcaption{Redshift evolution of the scalaron field.}
    \end{minipage}
    \caption{Redshift($z$) evolution of the density parameter. 
    The blue line shows the EoS parameter. The dashed orange line shows the EoS parameter for the scalaron field. The initial condition is $\Omega_r= 0.999, x=1.31\times10^{-11}, 
y= 1.31\times10^{-10}, \phi_0= 3$ at $z=5\times 10^5$.} 
    \label{Fig::Dark energy_large}
\end{figure}
We numerically evaluate the reheating process for a large decay width, $\Gamma/{\Lambda_{\rm Inf}}^{1/2}=10^{-3}$, and plot the results in Fig.~\ref{Fig::Reheating_large}.
We can find that the scalaron field slows down and freezes at $\phi\sim3 M_{\rm Pl}$.
We also evaluate the radiation and matter dominant era starting from $\phi_0\sim3 M_{\rm Pl}$ and show the results in Fig.~\ref{Fig::Dark energy_large}.
It is observed that the density parameter of the scalaron field, $\Omega_{\rm DE}$, increases slightly, but not enough to dominate the universe.
Thus, the results are not consistent with the current observation and the quintessetial inflation is not applicable.
This is because the potential is steep ($\epsilon_V\gg1$) in the positive $\phi$ region and the slow-roll scenario cannot be adopted in the dark energy era.
The quintessential inflation works well for a small decay width in the logarithmic model of Cartan $F(R)$ gravity.

\section{Conculusion}
\label{section::conculusion}
We have investigated the quintessential inflation in the logarithmic model of Cartan $F(R)$ gravity. 
For a small coupling, $\alpha$, the logarithmic model naturally has two significantly different scales. We tune the scales by two model parameters, $R_0$ and $\alpha$.
To realize the inflationary scale the parameter $R_0$ is fixed to be $R_0\propto \Lambda_{\rm Inf} e^{-\frac{1}{\alpha}}$. It means that the parameter $R_0$ is exponentially suppressed and we fix the parameter $\alpha$ to realize the scale close to the dark energy. Thus, we have succeeded in constructing the logarithmic model with the inflationary and dark energy scales. The potential has flat plateaus on each scale.

\begin{figure}
    \centering
    \includegraphics[scale=0.5]{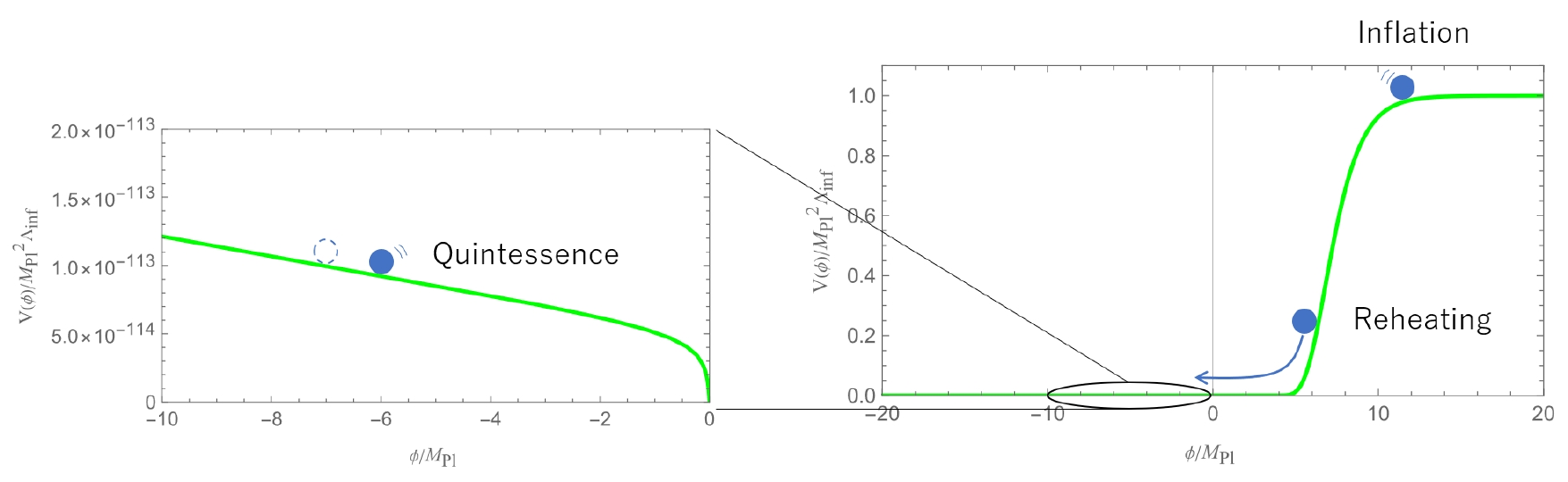}
    \caption{Overview of quintessential inflation by the logarithmic model in Cartan $F(R)$ gravity.}
    \label{fig:outline}
\end{figure}
We have evaluated analytically and numerically the process of time evolution of the universe.
The position of the scalaron field on the potential for each process is illustrated in Fig.~\ref{fig:outline}.
We have confirmed that the CMB fluctuations in the logarithmic model are consistent with the Planck 2018 results as in Fig.~\ref{Fig::CMBflu::Plancl}.
The reheating process is also completed through the kination process. 
After the hot Big Bang, the scalaron field freezes with potential energy at the dark energy scale during the radiation and matter dominance era and is hidden from the main player in the evolution of the universe.
The quintessence process is adapted for the negative $\phi$ region with the dark energy scale potential.
The redshift evolution of the density parameters and the scalaron field is numerically calculated
starting from the final state of reheating.
It is found that the potential energy of the scalaron field plays a role in the dark energy and causes the current accelerating expansion of the universe.

In this work, the reheating process has been evaluated with a constant decay width, $\Gamma$.
In Cartan form of general relativity, the torsion generates a four-fermion interaction for a spinor field called spin-spin interaction or Dirac-Heisenberg-Ivanenko-Hehl-Datta four-body fermi interaction~\cite{Hehl:1974cn,Kerlick:1975tr,Gasperini:1986mv,Hehl:1971qi,Boos:2016cey}.
The interaction between the spinor and the scalaron fields should be also considered in Cartan $F(R)$ gravity.
Thus, it is interesting to study the reheating process in the framework of Cartan $F(R)$ gravity with matter.
Here, we focus on the logarithmic model of Cartan $F(R)$ gravity. There is a possibility to construct Other models. These studies will give us a better understanding of the evolution of the universe in Cartan $F(R)$ gravity.
It is also interesting to discuss the verifiability of the quintessential inflation. In this direction,
we hope that the future gravitational wave observations \cite{Tashiro:2003qp,Riazuelo:2000fc,Gangopadhyay:2020bxn} will bring new and unknown information.


\section*{Acknowledgements}
For valuable discussions, the authors would like to thank H.~Sakamoto and N.~Yoshioka.
This work was supported by JST, the establishment of university fellowships towards the creation of science technology innovation, Grant Number JPMJFS2129.

\bibliography{ref}
\end{document}